\documentclass[ conference, letterpaper, 10pt, times]{IEEEtran}
\IEEEoverridecommandlockouts
\usepackage{cite}
\usepackage[T5]{fontenc}
\usepackage[utf8]{inputenc}
\usepackage{amsmath,amssymb,amsfonts}
\usepackage{graphicx}
\usepackage{textcomp}
\usepackage{xcolor}
\usepackage{graphicx}
\usepackage{tabularx}
\usepackage{stfloats}
\usepackage{multirow}
\usepackage{graphicx}   
\usepackage{subfig}     
\usepackage{url}
\DeclareRobustCommand*{\IEEEauthorrefmark}[1]{%
  \raisebox{0pt}[0pt][0pt]{\textsuperscript{\footnotesize #1}}%
  }
\usepackage{algorithm2e} 
\RestyleAlgo{ruled} 
\usepackage{array} 
\def\BibTeX{{\rm B\kern-.05em{\sc i\kern-.025em b}\kern-.08em
    T\kern-.1667em\lower.7ex\hbox{E}\kern-.125emX}}
    
\IEEEaftertitletext{\vspace{-2.5\baselineskip}} 
\begin{document}

\title{Scalable Impedance Identification of Diverse IBRs\\
via Cluster-Specialized Neural Networks}

\author{
    \IEEEauthorblockN{Quang Manh Hoang\IEEEauthorrefmark{1},
    Guilherme Vieira Hollweg\IEEEauthorrefmark{1},
    Bang Nguyen\IEEEauthorrefmark{3},
    Akhtar Hussain\IEEEauthorrefmark{2},
    Wencong Su\IEEEauthorrefmark{1,*}, 
    Van-Hai Bui\IEEEauthorrefmark{1,*}}
    \IEEEauthorblockA{\IEEEauthorrefmark{1}Department of Electrical and Computer Engineering, University of Michigan-Dearborn, USA}\IEEEauthorblockA{\IEEEauthorrefmark{2}Department of Electrical and Computer Engineering, Laval University, Canada}\IEEEauthorblockA{\IEEEauthorrefmark{3}National Renewable Energy Laboratory, Golden, CO, USA}
    \thanks{\textit{Corresponding authors: V.H. Bui (vhbui@umich.edu) and W. Su (wencong@umich.edu).}}
}

\maketitle

\begin{abstract}
Modern machine learning approaches typically identify the impedance of a single inverter-based resource (IBR) and assume similar impedance characteristics across devices. In modern power systems, however, IBRs will employ diverse control topologies and algorithms, leading to highly heterogeneous impedance behaviors. Training one model per IBR is inefficient and does not scale. This paper proposes a scalable impedance identification framework for diverse IBRs via cluster-specialized neural networks. First, the dataset is partitioned into multiple clusters with similar feature profiles using the K-means clustering method. Then, each cluster is assigned a specialized feed-forward neural network (FNN) tailored to its characteristics, improving both accuracy and computational efficiency. In deployment, only a small number of measurements are required to predict impedance over a wide range of operating points. The framework is validated on six IBRs with varying control bandwidths, control structures, and operating conditions, and further tested on a previously unseen IBR using only ten measurement points. The results demonstrate high accuracy in both the clustering and prediction stages, confirming the effectiveness and scalability of the proposed method.
\end{abstract}


\vspace{-0.5\baselineskip}
\section{Introduction}
Modern power grids are expected to integrate an increasing number of inverter-based resources, which makes the system more vulnerable and reduces the applicability of traditional control and stability analysis methods. One of the major challenges is the complex interaction among multiple IBRs due to their wide range of control bandwidths, topologies, and algorithms, together with the reduced inertia compared to traditional synchronous-generator (SGs)-based power systems. These interactions can lead to small-signal stability problems such as sub-synchronous oscillation, near-synchronous oscillation, and super-synchronous resonance \cite{shah2019impedance}.

In traditional power systems dominated by SGs, small-signal stability can be studied using white-box state-space models, because SGs have standardized structures and known parameters; therefore, eigenvalue analysis can be directly applied to investigate the system stability \cite{kundur2007power}. However, this approach is not applicable to inverter-dominated power systems, where the control structures and parameters of IBRs are often kept confidential by power converter manufacturers due to security and intellectual property considerations \cite{wu2023survey}. As a result, impedance-based stability analysis, which is well studied in the literature, has emerged as a powerful tool for assessing the small-signal stability of black-box models in the frequency domain \cite{shah2019impedance, zhou2022comprehensive, li2024machine}. In this method, the converter is treated as a black-box impedance model that relates voltage and current perturbations, typically obtained by injecting current or voltage at selected frequencies and measuring the corresponding responses \cite{francis2011algorithm}. Various impedance measurement methods based on frequency scanning have been developed for dc and ac converters.

However, measurement-based methods are known to be tedious and time-consuming, especially for large inverter-based power systems and wide frequency ranges. Even for a single IBR with detailed control structure, it may take several hours to obtain wide-frequency-range impedance characteristics at multiple operating points, and highly intermittent inverter-dominated systems require many such operating scenarios to be analyzed, which further increases the computational burden and complexity.

To overcome these drawbacks, several studies have focused on using data-driven methods to identify black-box IBR impedance models at multiple operating points \cite{11259993, zhang2020artificial, mohammed2024support, li2024machine}. In \cite{zhang2020artificial}, an artificial neural network (ANN) was used to identify the impedance model of a single grid-following inverter (GFLI) at different operating points; however, this model cannot predict impedance for converters with different control topologies, such as grid-forming inverters (GFMIs), without collecting a large amount of additional data. A related work in \cite{li2024machine} proposed a machine learning end-to-end framework based on transfer learning to identify GFLI impedance models at multiple operating points and with different control bandwidths. Nevertheless, the prediction accuracy of that model degrades significantly when applied to GFMIs, whose impedance characteristics differ from those of GFLIs. In \cite{mohammed2024support}, the authors used a support vector machine algorithm to predict the impedance of current-control GFLI, power-control GFLI, and virtual-synchronous-generator-based GFMI, but the model is retrained separately for each IBR rather than trained jointly. Notably, many existing machine-learning-based impedance modeling approaches struggle to generalize across highly heterogeneous IBR topologies. Although different IBRs may share identical input structures and operating variables, their fundamentally different dynamic behaviors lead to distinct impedance characteristics, which often results in degraded prediction accuracy when a single unified model is applied.

To address these limitations, this paper proposes a scalable impedance identification framework for diverse IBRs via cluster-specialized neural networks. The primary contribution lies in a system-level framework that addresses a practical and increasingly important challenge in large-scale power-electronic-dominated grids: scalable impedance identification under device heterogeneity. First, the dataset of six IBRs (three GFMIs and three GFLIs) with varying control bandwidths, control structures, and operating conditions is generated from analytical models. This dataset is then partitioned into multiple clusters with similar features using the K-means clustering method. Each cluster is subsequently assigned and trained with a specialized feed-forward neural network tailored to its feature profile, improving both accuracy and computational efficiency. In deployment, an unseen GFLI4 is tested using only ten measurement points at a single operating point. The results demonstrate high accuracy in both the clustering and prediction stages, confirming the effectiveness and scalability of the proposed method.
\vspace{-0.5\baselineskip}
\section{Proposed Framework}

\begin{figure*}[htbp]
\centerline{\includegraphics[scale = 0.63]{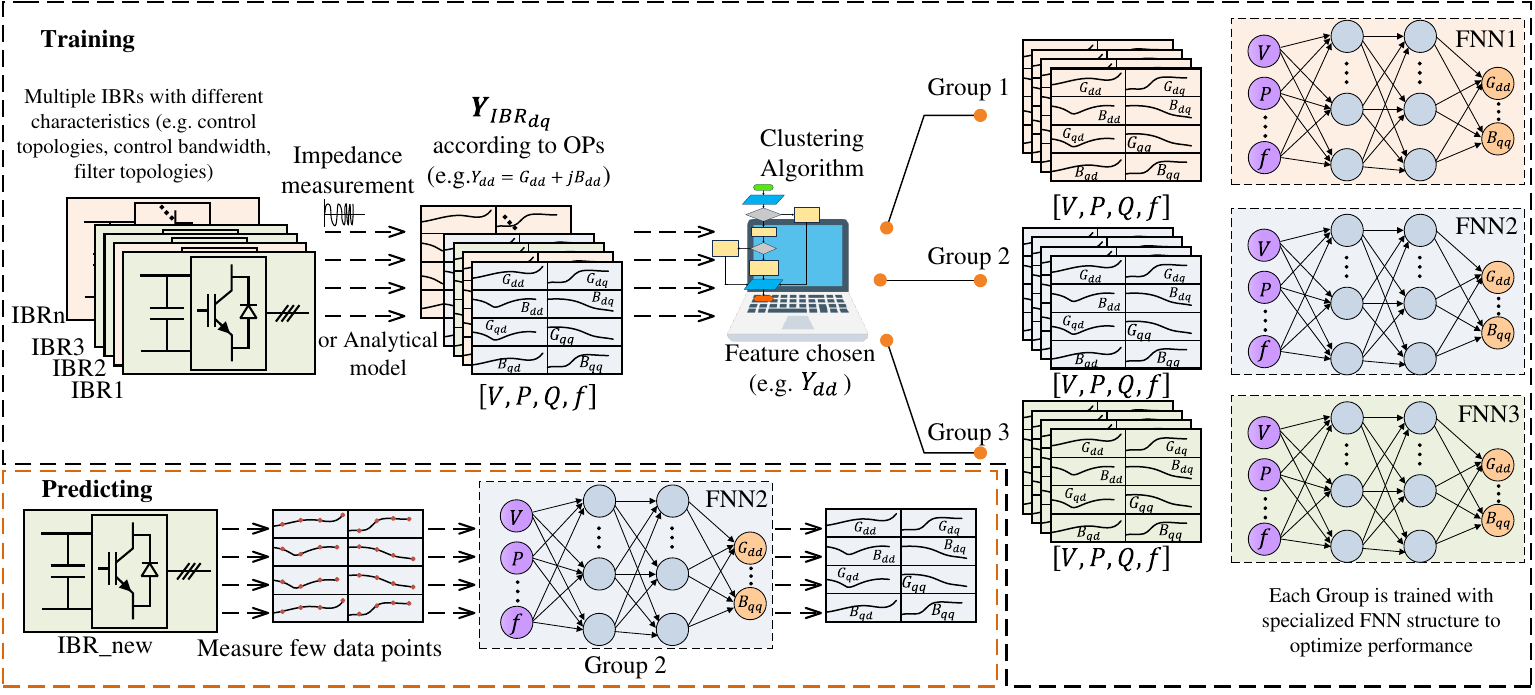}}
\caption{A clustering-driven framework for impedance identification in multi-IBR grids.}
\label{fig_framework}
\vspace{-1.5\baselineskip}
\end{figure*}

Figure~\ref{fig_framework} depicts the proposed framework, consisting of an offline phase (clustering and model training) and an online prediction phase. In the offline phase, impedance datasets for multiple IBRs are obtained either via measurement-based identification or from analytical models over a wide range of operating points (see Subsection~\ref{sec_dataacquis}). After assembling the dataset, one or more salient features (e.g., the $Y_{dd}$ component) are used for clustering (Subsection~\ref{sec_cluster}). Since the resulting clusters exhibit distinct characteristics, each cluster is assigned a dedicated feed-forward neural network architecture, as detailed in Subsection~\ref{sec_neuralnetwork}.

During deployment, the impedance of an unseen IBR can be identified from only a few measurements: first assign the device to its most likely cluster, then use the corresponding network to predict impedance across the broader operating range.
\vspace{-1\baselineskip}
\subsection{Data Acquisition}\label{sec_dataacquis}
\subsubsection{Generate dataset}

For model training, datasets can be obtained from offline or hardware-in-the-loop simulations; from field devices using measurement-based methods \cite{francis2011algorithm}; or from analytical models of IBRs \cite{li2024machine}. In this paper, the analytical approach is employed for both GFLI and GFMI impedance modeling, using typical configurations reported in the literature. The accuracy of this approach has been validated against measurement-based identification in \cite{xu2023small, zhou2022comprehensive}.

Analytical impedance models of both GFMIs and GFLIs have been extensively investigated \cite{wen2015analysis, zhou2022comprehensive, xu2023small, harnefors2007input}. Among the available techniques, the component-connection method is widely adopted due to its simplicity and ease of implementation \cite{harnefors2007input, zhou2022comprehensive, xu2023small}. In this method, linear subsystems, each described by a state-space model, are combined into a single overall system by stacking the state, input, and output vectors and assembling the corresponding coefficient matrices on the block diagonal. Interconnection relationships between subsystem inputs and outputs are then introduced, from which the composite system matrices $\boldsymbol{A_{\text{sys}}}, \boldsymbol{B_{\text{sys}}}, \boldsymbol{C_{\text{sys}}}, \boldsymbol{D_{\text{sys}}}$ are obtained by standard matrix algebra \cite{xu2023small}.

Once the overall state-space model of the IBR is available, the transfer function between the system inputs (containing $\boldsymbol{v_{{dq}}}$) and outputs (containing $\boldsymbol{i_{{dq}}}$) follows directly from control theory,
\vspace{-1\baselineskip}
\begin{equation}
    \boldsymbol{G(s)} = \boldsymbol{C_{\text{sys}}}(s\boldsymbol{I_{\text{sys}}} - \boldsymbol{A_{\text{sys}}})^{-1}\boldsymbol{B_{\text{sys}}} + \boldsymbol{D_{\text{sys}}},
\end{equation}
where $\boldsymbol{I_{\text{sys}}}$ is the identity matrix of the same dimension as $\boldsymbol{A_{\text{sys}}}$. The admittance in the $dq$ domain is then expressed as
\begin{equation}
    \mathbf{Y}_{dq}(s) = \frac{\mathbf{i}_{dq}(s)}{\mathbf{v}_{dq}(s)}.
\end{equation}

\subsubsection{Data structure}\label{sec_data}
Based on the analytical models of the GFMIs and GFLIs described in the previous section, six IBRs (three GFLIs and three GFMIs) are considered, with parameters listed in \cite{Manh2025Appendix}. The inputs represent the operating points of interest and the frequency range, defined as $(V, P, Q, f)$ in per-unit form, where $V \in [0.9, 1.1]$, $P \in [-1, 1]$, and $Q \in [-1, 1]$.

The training dataset is generated in MATLAB. A total of 100 frequency points are selected and evenly distributed on a logarithmic scale over $f \in [1, 200]$. The steps for $(V, P, Q)$ are chosen as $\{0.1,\, 0.5,\, 0.5\}$. The outputs consist of the conductance and susceptance values, that is, the real and imaginary parts of $Y_{dd}$, $Y_{dq}$, $Y_{qd}$, and $Y_{qq}$. Therefore, each sample has a dimension of $\mathbb{R}^{1 \times 12}$. Some generated operating points exceed the rated power limit (apparent power greater than 1 pu) and are discarded, resulting in a total of 3900 samples per IBR.

The testing dataset is generated using the same principles, but with finer resolution: 200 frequency points per operating point and steps for $(V, P, Q)$ equal to $\{0.1,\, 0.2,\, 0.2\}$. Operating points that exceed the rated power are similarly removed, yielding a total of 48{,}600 testing samples.

Since the training dataset aggregates samples from all six IBRs, training a single FNN for all devices is infeasible. Although the inputs share an identical structure and values, each IBR produces different output characteristics. This mismatch leads to poor prediction accuracy and explains why existing studies do not report impedance prediction for multiple heterogeneous IBRs using a single model. Therefore, in this research, the dataset must be partitioned into subgroups with similar features before training, as presented in the next section using a clustering algorithm.

\subsection{Clustering Module}\label{sec_cluster}
After collecting the data from all six IBRs, the samples must be clustered into subgroups for training purposes. If all input features are used directly, the strong similarity in input structure can lead to poor clustering performance. Therefore, only a representative feature is selected for the clustering step. In this paper, the magnitude of the $dd$-admittance, $|Y_{dd}|$, is chosen as a minimal yet effective feature for scalability, and the K-means clustering algorithm is employed \cite{likas2003global}. Although a single feature is used in this work, the proposed framework itself does not restrict the choice of clustering features and can be readily extended to multi-feature or weighted-feature configurations when needed.

\begin{figure}[t!]
\centerline{\includegraphics[scale=0.5]{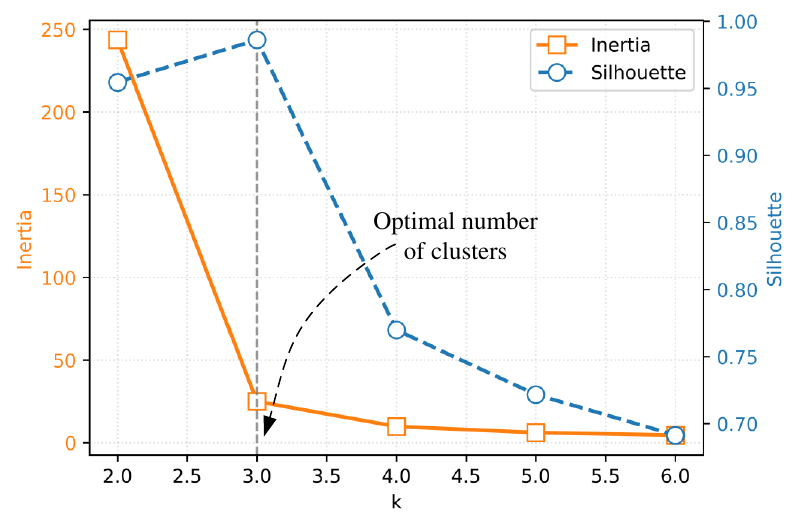}}
\caption{Inertia and silhouette score for each number of clusters.}
\label{fig_Iner}
\vspace{-1\baselineskip}
\end{figure}

The next challenge is to determine an appropriate number of clusters, $K_{\text{means}}$. To address this, the K-means clustering algorithm is executed with different numbers of clusters ranging from 2 to 6. Two criteria, inertia and silhouette score, are used to evaluate the quality of the resulting partitions. Inertia quantifies the internal compactness of clusters by measuring the sum of squared distances between data points and their corresponding centroids; lower values indicate tighter and more coherent cluster structures. The silhouette score captures both the cohesion of a point within its assigned cluster and its separation from other clusters; higher average values reflect well-defined cluster boundaries and overall superior clustering quality \cite{likas2003global}. The variation of these two indices with respect to the number of clusters is shown in Fig.~\ref{fig_Iner}. By computing the average rank over both criteria, three clusters are selected as the optimal choice for this dataset.

\begin{figure}[t!]
\centerline{\includegraphics[scale=0.6]{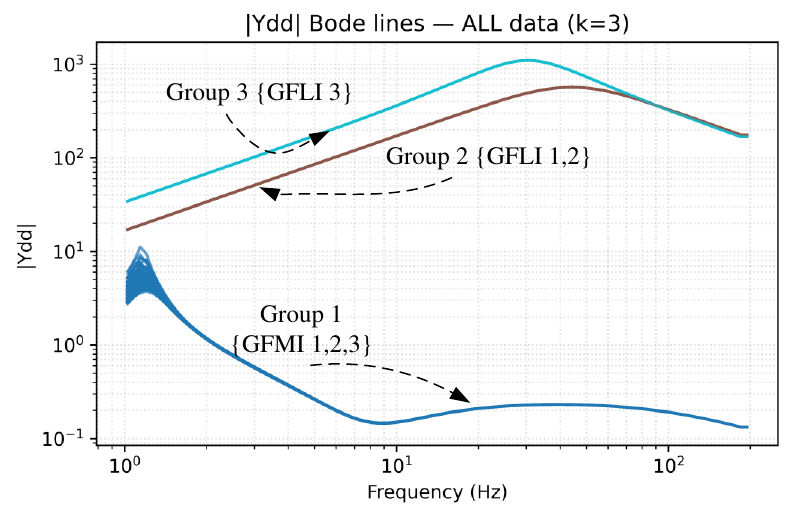}}
\caption{Magnitude of the $dd$-admittance for all operating points grouped by cluster.}
\label{fig_Clus}
\vspace{-1.5\baselineskip}
\end{figure}

After running the K-means clustering algorithm with three clusters, the magnitudes of the $dd$-admittance for all operating points are plotted in Fig.~\ref{fig_Clus}. As shown, Group~1 contains GFMI1, GFMI2, and GFMI3 with different power control bandwidths, leading to larger variance and making this group more challenging to model accurately. Group~2 contains GFLI1 and GFLI2 with very small variance, while Group~3 contains GFLI3. It can be observed that each group exhibits distinct characteristics, especially Group~1. Hence, it is essential to design specialized FNN models for each group. These aspects are further discussed in the next section.
\vspace{-0.5\baselineskip}

\subsection{Cluster-Specialized Neural Networks}\label{sec_neuralnetwork}
A state-of-the-art feed-forward neural network (FNN) architecture is employed to train three separate models, one for each cluster, using common hyperparameter settings summarized in Table~\ref{tb_FNNParams} \cite{bebis2002feed}. The networks are trained using the mean squared error (MSE) loss and the Adam optimizer with mini-batch training. The only difference between the three models is the number of neurons in the hidden layers. In particular, the model for Group~1 uses fewer neurons per layer to reduce overfitting, because Group~1 exhibits greater variability in the data, especially in the low-frequency range, whereas Groups~2 and~3 are more homogeneous. The values of the hyperparameters are selected by manual trial-and-error.

\begin{table}[!h]
\caption{FNN model and training settings}\label{tb_FNNParams}
\centering
\begin{tabular}{>{\centering\arraybackslash}p{1.3in} >{\centering\arraybackslash}p{0.5in}>{\centering\arraybackslash}p{0.5in} >{\centering\arraybackslash}p{0.5in}}
\hline\hline
 & FNN 1 & FNN 2 & FNN 3 \\ \hline
Maximum iterations   &  1200 & 1200 & 1200   \\ 
Activation function  &  ReLU & ReLU & ReLU \\
Solver               &  Adam & Adam & Adam  \\
Hidden layers        &  (4, 8) & (32, 32) & (32, 32)  \\
\hline
\hline
\end{tabular}
\vspace{-1\baselineskip}
\end{table}

A related study has employed the Optuna hyperparameter-optimization framework integrated with TensorFlow \cite{li2024machine}. However, automated hyperparameter tuning is not the primary focus of this work. In the present implementation, the datasets are preprocessed using the \texttt{scikit-learn} library, and the networks are trained on a personal computer equipped with an Intel 13th Gen Core i5-13500HX processor.

\begin{figure}[!h]
\centering
\subfloat[Conductance on testing.\label{fig_Conduc}]{
    \includegraphics[width=0.48\linewidth]{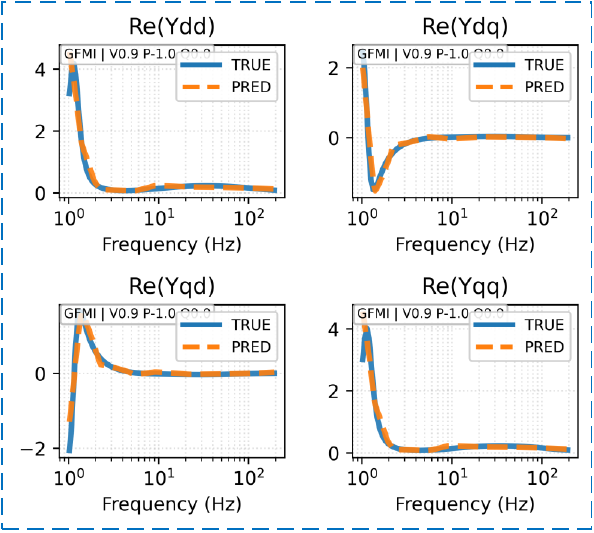}
}
\subfloat[Susceptance on testing.\label{fig_Suscep}]{
    \includegraphics[width=0.48\linewidth]{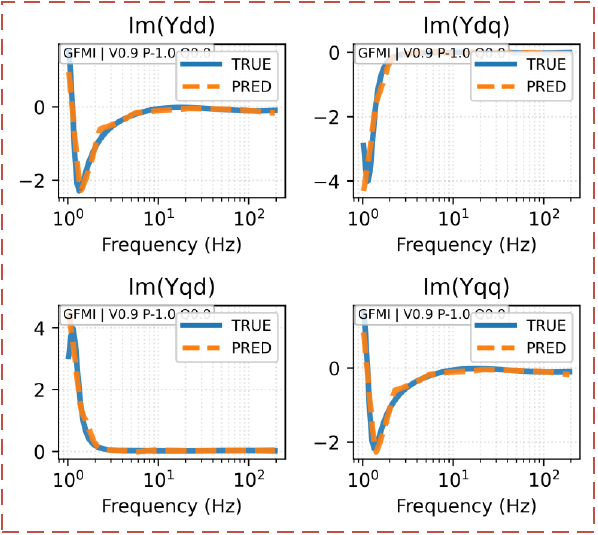}
}
\caption{Testing performance of the trained FNN model for Group~1.}
\label{fig_Test}
\vspace{-1.5\baselineskip}
\end{figure}

To illustrate the effectiveness of the proposed framework, detailed testing results for Group 1 are presented, as shown in Fig.~\ref{fig_Test}. Figure~\ref{fig_Conduc} presents the true and predicted conductance for the four admittance components of $\mathbf{Y}_{dq}$ at the operating point $V = 0.9$~pu, $P = 1$~pu, and $Q = 0$. Figure~\ref{fig_Suscep} shows the corresponding susceptance at the same operating point. As mentioned in Section~\ref{sec_dataacquis}, the testing dataset has finer resolution than the training dataset, which provides a stringent evaluation of the generalization capability of the FNN. As can be seen in these figures, the model predicts the admittance with high accuracy across the studied frequency range. 

For further validation, an unseen grid-following inverter, denoted GFLI4, with different PLL control parameters is used to evaluate the effectiveness of the proposed framework, and the corresponding results are discussed in the next subsection.
\vspace{-1.5\baselineskip}

\section{Case Study and Validation}
\begin{figure*}[htbp]
    \centering
    \subfloat[True (analytical) conductance of GFLI4.\label{fig_Gener}]{
        \includegraphics[width=0.8\linewidth]{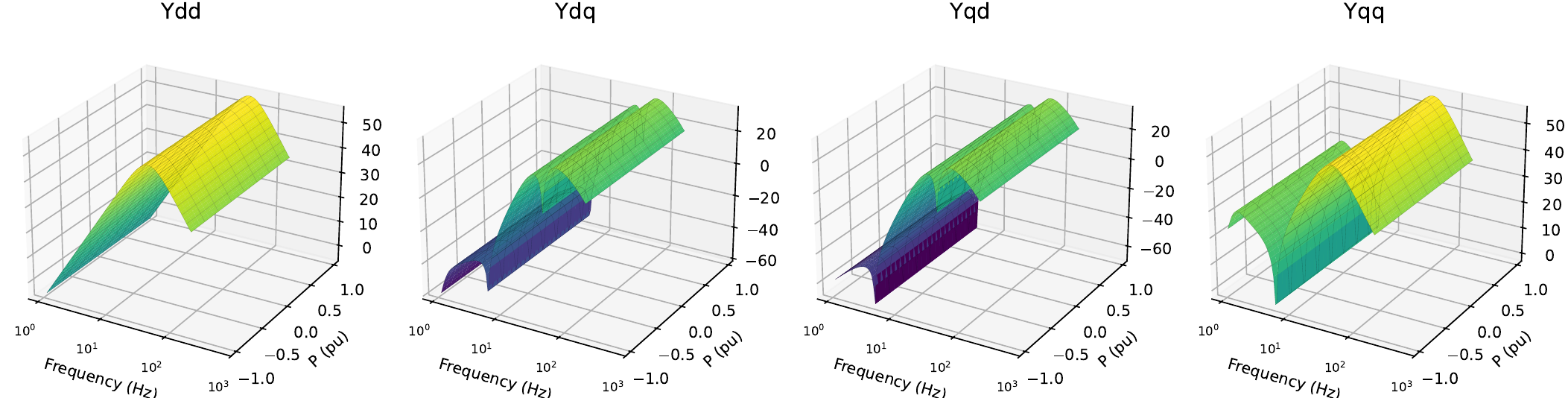}
    }\vspace{0.5\baselineskip}
    \subfloat[Predicted conductance of GFLI4.\label{fig_Pred}]{
        \includegraphics[width=0.8\linewidth]{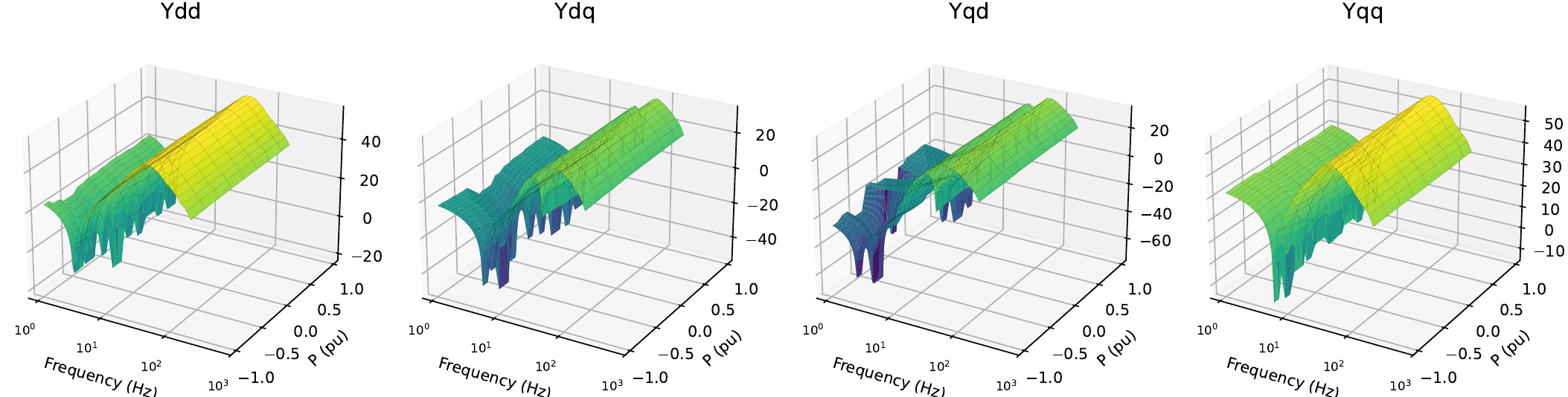}
    }\vspace{0.5\baselineskip}
    \subfloat[Prediction error (predicted - true).\label{fig_Error}]{
        \includegraphics[width=0.8\linewidth]{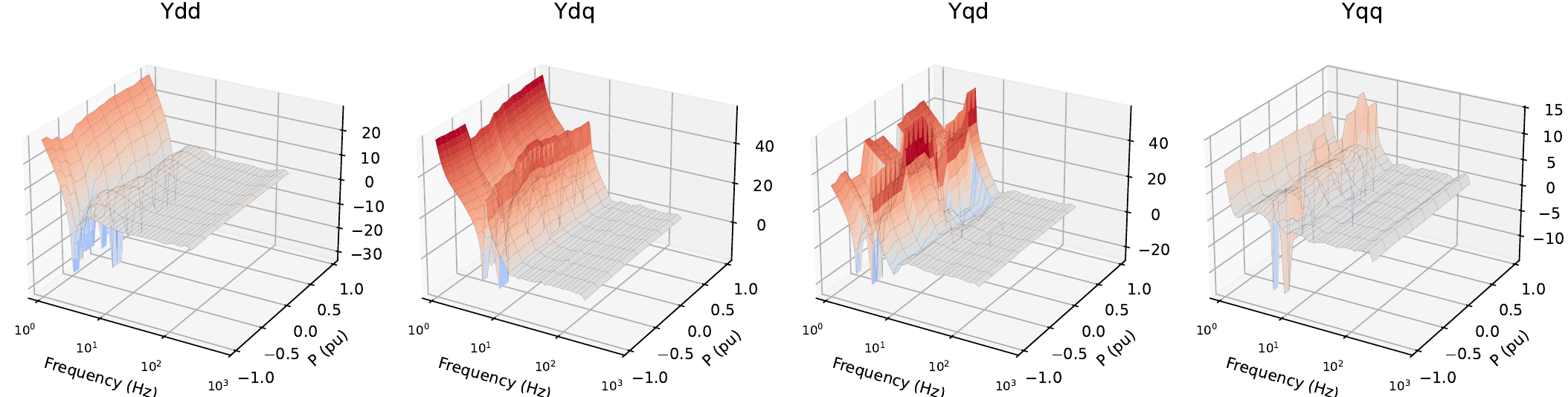}
    }
    \caption{True, predicted, and corresponding errors of the conductance components of the GFLI4 admittance over the studied operating points.}
    \label{fig_Validate}
    
\end{figure*}

In this section, an unseen inverter, denoted GFLI4, with parameters identical to those of GFLI1 in previous section, except $K_{ppll} = 7.9$ and $K_{ipll} = 78.9$, is used to validate the proposed framework under the assumption that only ten frequency points at a single operating point ($V = 1$~pu, $P = 1$~pu, $Q = 0$~pu) are available. First, the magnitude of the $dd$-admittance is computed and normalized in the same way as in the clustering stage. During training, the centroids of the three clusters, $\mu^{(1)},\ \mu^{(2)},\ \mu^{(3)}$, have already been calculated and stored. Hence, whenever a new IBR needs to be assigned to a cluster, the Euclidean distance between its $|Y_{dd}|$ profile and each centroid is computed as
\begin{equation}
d_k = \left\lVert z - \mu^{(k)} \right\rVert_2,
    \qquad k = 1, 2, 3,
\end{equation}
where $z$ denotes the normalized $|Y_{dd}|$ vector of the new IBR. The new device is then assigned to the cluster with the smallest distance.

For GFLI4, the distances to Groups~1, 2, and 3 are $\{d_1, d_2, d_3\} = \{18.7,\ \boldsymbol{1.17},\ 3.9\}$, so the model associated with Group~2 (FNN2) is selected for predicting the admittance over a wide range of operating points and frequencies. The prediction results are shown in Fig.~\ref{fig_Validate}, where the voltage and reactive power are fixed at 1~pu and 0~pu, respectively, for better visualization. The active power is varied from $-1$~pu to $1$~pu with a step of 0.2, and 200 frequency points are considered for each operating point.

As can be seen from these figures, FNN2 accurately predicts the admittance of the new GFLI across most of the studied frequency range. The largest discrepancies appear at low frequencies, which is expected because the PLL dynamics typically dominate in this region and are more sensitive to parameter variations. Nevertheless, the overall agreement confirms that the cluster-specialized model can generalize well to a previously unseen IBR using only a small number of measurement points.

In practice, when a new IBR is integrated into the system, the selected cluster model can be further fine-tuned with a few additional measurements from that device, especially in the low-frequency range, to improve accuracy if required. This illustrates a practical workflow in which the proposed framework provides a good initial impedance model with minimal data, and optional fine-tuning can be applied to meet stricter accuracy demands. This case study illustrates the few-shot capability of the proposed cluster-specialized framework for modeling heterogeneous IBRs.

\section{Conclusion}
This paper proposed a clustering-based machine learning framework for impedance identification of diverse inverter-based resources. Impedance datasets for three grid-forming and three grid-following inverters with different control parameters were generated from analytical models and partitioned into three clusters, and a specialized feed-forward neural network was trained for each cluster, reducing computational resources and model complexity compared with assigning one model per IBR. The framework was further validated on an unseen grid-following inverter using only ten measurement points at a single operating point, showing high prediction accuracy over most of the frequency range, with larger errors concentrated at low frequencies where phase-locked loop dynamics dominate. Although the proposed framework addresses the scalability challenge, it still relies on centralized data aggregation, which may introduce cybersecurity and privacy concerns; future work will investigate secure and privacy-preserving implementations and extend the approach to experimental datasets and larger IBR fleets.

\section*{Acknowledgment}

This work was supported in part by the U.S. National Science Foundation (NSF) under Award 2509993.

\bibliographystyle{ieeetr}
\bibliography{Ref}

\end{document}